\documentstyle[epsfig]{aipproc}

\begin{document}
\title{Recent Progress in Understanding Nucleosynthesis via Rapid\\
       Neutron Capture}

\author{Yong-Zhong Qian}
\address{School of Physics and Astronomy, University of Minnesota,
Minneapolis, MN 55455}

\maketitle

\begin{abstract}
I discuss the recent progress in our understanding of nucleosynthesis
via rapid neutron capture, the $r$-process, based on meteoritic data
for the early solar system and observations of stars at low 
metallicities. At present, all data require that there be two distinct
kinds of $r$-process events and suggest that supernovae are associated
with these events. The diversity of supernova sources for the 
$r$-process may depend on whether a neutron star or black hole is
formed in an individual supernova. This dependence, if substantiated
by future observations discussed here, has important implications
for properties of nuclear matter.
\end{abstract}

\section*{Introduction}
The grand scheme for production of various elements was set down 
more than forty years ago \cite{bbfh,al}. 
Within this grand scheme, approximately half of
the heavy elements with mass numbers $A>100$ in the solar system
were produced via rapid neutron capture, the $r$-process. A crude
picture for the $r$-process is as follows. One starts with some seed
nuclei and lots of neutrons. The seed nuclei then rapidly capture
these neutrons to make very neutron-rich unstable progenitor nuclei.
After neutron capture stops, the progenitor nuclei successively
$\beta$-decay towards stability and become the $r$-process nuclei
observed in nature. For a given species of seed nuclei with mass number
$A_s$, the group of nuclei produced by the $r$-process are determined 
by the number of neutrons per seed nucleus, or the neutron-to-seed ratio 
$n/s$, at the beginning of the $r$-process. The average mass number of
the produced $r$-process nuclei is $\langle A\rangle=A_s+n/s$.

There are two prominent peaks at $A=130$ and 195, respectively,
in the solar $r$-process abundance pattern. In order to produce
the peak at $A=130$, we need $n/s\sim 40$ if we start from seed
nuclei with $A_s\sim 90$. By comparison, if we start with the
same seed nuclei but a higher $n/s\sim 90$, the $r$-process dominantly
produces nuclei with $A>130$, including the peak at $A=195$.
In general, an astrophysical event would eject, for example, a certain
amount of $r$-process material with $n/s\sim 40$ plus some other amount
with $n/s\sim 90$. The ratio of these two amounts then determines the
overall $r$-process abundance pattern produced by this event. A natural
question is whether every $r$-process event produces the same abundance
pattern or there should be distinct kinds of events producing very
different $r$-process abundance patterns. In other words, we would like
to know whether the solar $r$-process abundance pattern is produced by
every $r$-process event or just reflects a mixture of different patterns
produced by distinct kinds of events. In either case, we also would like 
to know which astrophysical objects are associated with the $r$-process.

Here I discuss the progress that we have made recently in answering the
above questions. The answers to these questions have important 
implications for properties of neutrinos, nuclei far from stability, and
nuclear matter. I discuss the implications for properties of 
nuclear matter in particular.

\section*{Meteoritic data and diverse supernova sources for the 
$r$-process}
The meteoritic data on the inventory of radioactive nuclei in the 
interstellar medium (ISM) at the time of solar system formation play an
essential role in our understanding of the $r$-process. Because a 
radioactive species decays after it is produced, a finite abundance
ratio of a radioactive species to a stable one in the ISM has to be 
maintained by a series of production events. Consequently, from the 
measured abundance ratio of a radioactive species to a stable one, we
can infer how frequently the radioactive species was replenished in the
ISM. We have data on two radioactive nuclei: $^{129}$I 
(e.g., \cite{i129}) and $^{182}$Hf (e.g., \cite{hf182}), 
which are below and above $A=130$, respectively.
If the data indicate that these two species were injected into the ISM
at very different frequencies, then the $r$-process nuclei below and 
above $A=130$ must be produced by distinct kinds of events. It was found
that over the Galactic history of $\approx 10^{10}$~yr before solar 
system formation, $^{182}$Hf injection occurred at a high frequency 
$f_{\cal{H}}\sim (10^7\ {\rm yr})^{-1}$, while $^{129}$I injection 
occurred at a low frequency $f_{\cal{L}}\sim (10^8\ {\rm yr})^{-1}$
\cite{wbg,qw}. Therefore, these are distinct kinds of events.

The frequencies of $^{129}$I and $^{182}$Hf injection into the ISM
inferred from meteoritic data can be naturally explained by associating
supernovae with the $r$-process. A supernova remnant expands to a final
radius of $\sim 100$~pc over a period ($\sim 10^6$~yr)
much shorter than the lifetime ($\sim 10^7$~yr) of $^{129}$I or 
$^{182}$Hf. Therefore, if we consider a spherical region of 
$\sim 100$~pc in radius surrounding an average point in the Galaxy,
any supernova within this region can inject fresh radioactive $^{129}$I
or $^{182}$Hf to this point after it is produced by the supernova.
For a supernova frequency of $\sim (30\ {\rm yr})^{-1}$ over the
Galactic volume of $\sim 10^3\ {\rm kpc}^3$, the corresponding
frequency in this spherical region is $\sim (10^7\ {\rm yr})^{-1}$.
Therefore, the meteoritic data on $^{129}$I and $^{182}$Hf can be
explained if we associate the most common supernovae with the 
$r$-process events producing $^{182}$Hf and a rarer kind with those
producing $^{129}$I \cite{wbg,qw,qvw1}.

Because $^{129}$I is produced together with nuclei at $A\leq 130$
and $^{182}$Hf with those at $A>130$, we conclude
that the overall solar $r$-process abundance pattern is composed of
two basic templates characteristic of two distinct kinds of
$r$-process events. These are referred to as the ``$\cal{H}$'' and 
``$\cal{L}$'' events hereafter, where ``$\cal{H}$'' stands for the 
``high'' frequency events responsible for ``heavy'' $r$-process nuclei 
with $A>130$ including $^{182}$Hf and ``$\cal{L}$'' for the 
``low'' frequency events responsible for ``light'' $r$-process nuclei 
with $A\leq 130$ including $^{129}$I. An average ISM is enriched in 
$r$-process elements at a frequency 
$f_{\cal{H}}\sim (10^7\ {\rm yr})^{-1}$ by the $\cal{H}$ events and
at a frequency $f_{\cal{L}}\sim (10^8\ {\rm yr})^{-1}$ by the 
$\cal{L}$ events.

\section*{Abundances of $r$-process elements in metal-poor stars}
The influence of the $\cal{H}$ and $\cal{L}$ events is best preserved 
in metal-poor stars formed very early in the Galaxy when only a small 
number of supernovae had contributed to the $r$-process and ``metal''
abundances in these stars. 
The typical heavy $r$-process elements observed in these stars are Ba 
and Eu, and the typical light $r$-process elements observed are Pd, Ag, 
and Cd. The observational data are usually given in the spectroscopic 
notation, e.g., $\log\epsilon({\rm Eu})\equiv\log({\rm Eu/H})+12$ for 
Eu, where Eu/H is the number abundance ratio of Eu to hydrogen observed 
in a star. A typical metal is Fe, and the ``metallicity'' is
defined as [Fe/H]~$\equiv\log({\rm Fe/H})-\log({\rm Fe/H})_{\odot}$,
where $({\rm Fe/H})_{\odot}$ is the Fe/H ratio in the sun.
As the overall abundance of hydrogen has not changed 
significantly over the history of the universe, hydrogen is a good
reference element for considering chemical enrichment of the ISM.

Over the period of $\approx 10^{10}$~yr before solar system formation,
an average ISM was enriched in the heavy $r$-process elements by
$\sim 10^3$ $\cal{H}$ events and in the light ones by $\sim 10^2$ 
$\cal{L}$ events. The $r$-process composition of the ISM at the time 
of solar system formation is reflected by the corresponding solar 
abundances. Consequently, the $r$-process abundances resulting from a 
single $\cal{H}$ or $\cal{L}$ event (quantities with the subscript
``$\cal{H}$'' or ``$\cal{L}$'') can be predicted directly from the 
solar system data (quantities with the subscript ``$\odot,r$'') 
\cite{qw,wq}. For example,
we have ${\rm (Eu/H)}_{\odot,r}\sim 10^3{\rm (Eu/H)}_{\cal{H}}$, and hence 
$\log\epsilon_{\cal{H}}({\rm Eu})\sim
\log\epsilon_{\odot,r}({\rm Eu})-3\approx-2.5$.
Likewise, we have ${\rm (Ag/H)}_{\odot,r}\sim 10^2{\rm (Ag/H)}_{\cal{L}}$,
and hence 
$\log\epsilon_{\cal{L}}({\rm Ag})\sim
\log\epsilon_{\odot,r}({\rm Ag})-2\approx-0.8$.

\begin{figure}[b!]
\centerline{\epsfig{file=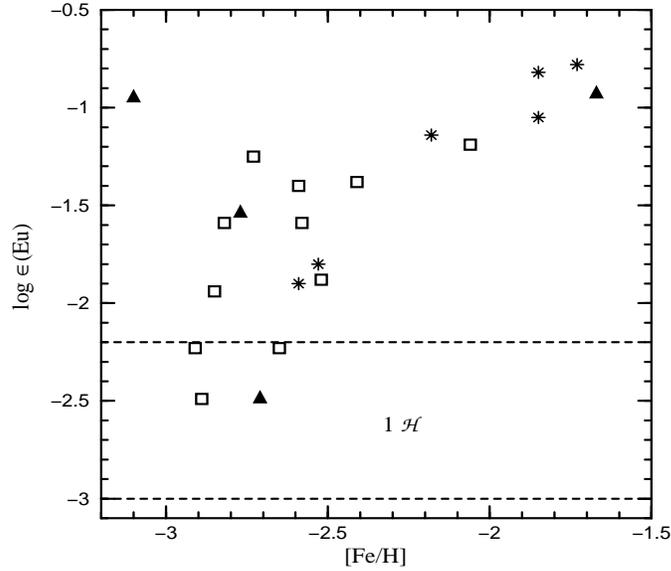,height=3in,width=3.5in}}
\vspace{10pt}
\caption{Europium data for metal-poor stars (asterisks: 
\protect\cite{gratton}, squares: \protect\cite{mcwill},
triangles: \protect\cite{sneden1,sneden2})
compared with the Eu abundance
resulting from a single $\cal{H}$ event (the band labeled ``1 $\cal{H}$'') 
predicted in \protect\cite{qw,wq}.}
\label{eu}
\end{figure}

\begin{figure}[b!]
\centerline{\epsfig{file=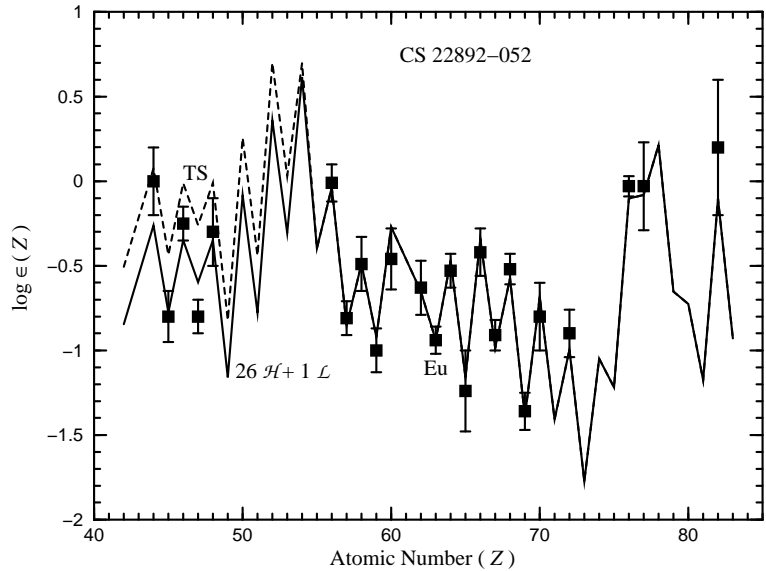,height=4in,width=3in,angle=270}}
\vspace{10pt}
\caption{Comparison of data for CS 22892--052 with (1) the solar $r$-process
abundance pattern translated to match the Eu data (dashed curve), and
(2) the result from a mixture of two distinct kinds of $r$-process events 
(solid curve).}
\label{cs}
\end{figure}

Figure \ref{eu} shows the Eu data for many metal-poor stars
\cite{gratton,mcwill,sneden1,sneden2}. 
The very low metallicities of these stars
indicate that they were formed very early in the Galaxy. The band
labeled ``1 $\cal{H}$'' corresponds to the Eu abundance resulting
from a single $\cal{H}$ event predicted by the meteoritic data on
$^{129}$I and $^{182}$Hf as well as the solar $r$-process abundances 
of stable nuclei \cite{qw,wq}. If we pick the centroid of this band,
the observed Eu abundances can be explained by contributions from
$\sim 1$--30 $\cal{H}$ events, quite consistent with the corresponding
low metallicities.

Figure \ref{cs} shows the remarkable data on a number of $r$-process
elements for one of the stars shown in Figure \ref{eu}, CS 22892--052
\cite{sneden3}. The dashed curve labeled ``TS''
is the solar $r$-process abundance
pattern translated to match the data on the heavy $r$-process elements.
Clearly, this curve cannot match the data on the light
$r$-process elements. Therefore, the conclusion from the meteoritic
data that there should be two distinct kinds of $r$-process events
is independently confirmed by stellar observations at low 
metallicities. The Eu abundance in CS 22892--052 can be explained by
$\sim 30$ $\cal{H}$ events. However, with a frequency ratio of
$\sim 10:1$ between $\cal{H}$ and $\cal{L}$ events, it is very likely
that one $\cal{L}$ event also occurred during a period over which
$\sim 30$ $\cal{H}$ events took place. Indeed, a mixture of 26 
$\cal{H}$ events and 1 $\cal{L}$ event (solid curve in Fig. \ref{cs})
can explain all the data rather well \cite{qw}.

\section*{Neutron star/black hole formation and supernova $r$-process 
nucleosynthesis}
So far, the questions raised in the introduction have been answered. 
It is found that there are distinct kinds of $r$-process events and
that they are associated with supernovae. This leads to a new
question: what is causing the difference between these supernova
$r$-process events? The answer to the new question may be obtained by
further studying $r$-process enrichment of the Galaxy by supernovae.
The present Galactic inventory of either the light $r$-process nuclei
($100\lesssim A\leq 130$) or the heavy ones ($A>130$) is
$\sim 4\times 10^3\,M_{\odot}$. Assuming a frequency of 
$\sim (30\ {\rm yr})^{-1}$ over the whole Galaxy in its history of 
$\approx 10^{10}$~yr for the supernovae associated with $\cal{H}$ 
events and a $\sim 10$ times less frequency for those associated with 
$\cal{L}$ events, we find that in order to account for the present 
Galactic $r$-process inventory, each supernova has to eject only 
$\sim 10^{-5}\,M_{\odot}$ ($\cal{H}$ event) to $\sim 10^{-4}\,M_{\odot}$ 
($\cal{L}$ event) of $r$-process material. By comparison, the total
amount of ejecta from a supernova is $\sim 10\,M_{\odot}$ and the mass
of the neutron star produced in a supernova is $\sim 1\,M_{\odot}$.
Despite the striking difference from these comparison mass scales, 
$\sim (10^{-5}$--$10^{-4})\,M_{\odot}$ of material can be naturally 
ejected by the neutrinos emitted in a supernova.

A supernova occurs when the core of a massive star at the exhaustion
of its nuclear fuel collapses into a compact neutron star. The neutron
star has a final radius of $\sim 10$~km, and a gravitational binding
energy of $\sim 10^{53}$~erg is to be released. Due to the high 
temperatures and densities encountered during the collapse, the most
efficient way to release this energy is to emit all three flavors of
neutrinos and 
antineutrinos mainly through electron-positron pair annihilation.
In fact, because of the intense scatterings on neutrons and protons
common to all neutrino species, even neutrinos have to diffuse out of
the neutron star on a timescale of $\sim 10$~s (as confirmed by the
detection of neutrinos from SN 1987A). So the average neutrino 
luminosity is $\sim 10^{51}$~erg s$^{-1}$ per species.

A few seconds after the core collapse and the subsequent supernova
explosion, we have a hot neutron star near the center of the supernova.
The neutron star is still cooling by emitting neutrinos. The shock
wave which makes the supernova explosion is far away from the neutron
star. On its way out to make the explosion, the shock wave has cleared
away almost all the material above the neutron star, leaving behind
only a thin atmosphere. Close to the neutron star, the temperature is
several MeV and the atmosphere is essentially dissociated into neutrons
and protons. As the neutrinos emitted by the neutron star free-stream
through this atmosphere, some of the $\nu_e$ and $\bar\nu_e$ are
captured by the neutrons and protons and their energy is deposited
in the atmosphere. In other words, the atmosphere is heated by the
neutrinos. As a result, it expands away from the neutron star and
eventually develops into a mass outflow --- a neutrino-driven ``wind''
\cite{duncan}.

Because neutrino heating is driving the mass ejection, the fraction
of neutrino luminosity absorbed by the wind material determines the
rate at which it is being lifted out of the neutron star
gravitational potential. As neutrinos interact weakly,
the heating rate is small. On the other hand, the neutron
star is a compact object and has a deep gravitational potential.
Consequently, we expect that the mass ejection rate is small.
Indeed, the typical mass ejection rate in the wind was found to be
$\sim 10^{-5}\,M_{\odot}\ {\rm s}^{-1}$ \cite{qwoo}. So provided that
$r$-process nuclei are produced in the neutrino-driven wind,
the wind has to last $\sim 1$~s in an $\cal{H}$ event and 
$\sim 10$~s in an $\cal{L}$ event in order to account for the present 
Galactic $r$-process inventory. Note that
neutrino emission from a stable neutron star, and hence the 
corresponding neutrino-driven wind, last $\sim 10$~s. In order for
a wind to last only $\sim 1$~s, neutrino emission has to be
terminated $\sim 1$~s after the supernova explosion by transformation
of the neutron star into a black hole \cite{qvw1}. Therefore, the
difference between the $\cal{H}$ and $\cal{L}$ events may depend
on whether a black hole ($\cal{H}$ event) or neutron star ($\cal{L}$ 
event) is formed in an individual supernova.

\section*{Conclusions}
In summary, meteoritic data on $^{129}$I and $^{182}$Hf require
two distinct kinds of $r$-process events: the high frequency $\cal{H}$
events responsible for heavy $r$-process nuclei ($A>130$) and the
low frequency $\cal{L}$ events responsible for light ones 
($100\lesssim A\leq 130$). The meteoritic data also suggest that 
supernovae are associated with $\cal{H}$ and $\cal{L}$ events. 
These conclusions are either confirmed by or at the very least 
consistent with observations of $r$-process abundances in metal-poor 
stars.

If future studies can show that $r$-process nuclei are produced in
the neutrino-driven wind in a supernova (see e.g., 
\cite{qwoo,meyer1,janka1,janka2,woo,hoff,hax1,hax2,meyer2,fuller,qian}
for the current unsatisfactory status), then the amount 
of material required to account for the present Galactic $r$-process 
inventory can be adequately provided by the wind. The factor 
of $\sim 10$ difference in the amount of $r$-process ejecta between the 
$\cal{H}$ and $\cal{L}$ events calls for a similar difference in the 
duration of neutrino emission, and hence that of the neutrino-driven 
wind, between the corresponding supernovae. In turn, the difference in 
neutrino emission may require transformation of the initial neutron star 
into a black hole $\sim 1$~s after the supernova explosion in an 
$\cal{H}$ event and long term stability of the neutron star in an 
$\cal{L}$ event. Consequently, the diversity of supernova sources for 
the $r$-process may have profound implications for properties of nuclear 
matter inside the initial neutron star produced in a supernova.

Two possible observational tests for the association of neutron 
star/black hole formation with supernova $r$-process nucleosynthesis 
have been proposed \cite{qian,qvw2,qvw3}. One test relies on the
occurrence of supernovae in binaries consisting of a massive star and
a low mass star. Some binaries would survive the supernova explosion of 
the massive star and become new systems with a neutron star or 
black hole orbiting around the low mass star. Furthermore, the surface
of the low mass star would be contaminated by the $r$-process ejecta 
from the supernova. Therefore, we can test black hole and neutron star
formation in $\cal{H}$ and $\cal{L}$ events, respectively,
by looking for $r$-process
abundance anomalies on the surface of the binary companion to a neutron
star or black hole \cite{qian}. This approach is quite promising as 
large overabundances of O, Mg, Si, and S ejected in supernovae have 
been observed in the binary companion to a black hole 
\cite{israel}.

\section*{Acknowledgments}
This work was supported in part by the Department of Energy under
grant DE-FG02-87ER40328.


\begin{references}
\bibitem{bbfh}Burbidge, E. M., et al.,
{\it Rev. Mod. Phys.} {\bf 29}, 547 (1957).
\bibitem{al}Cameron, A. G. W., {\it Pub. Astron. Soc. Pacific} {\bf 69}, 
201 (1957).
\bibitem{i129}Reynolds, J. H., {\it Phys. Rev. Lett.} {\bf 4}, 8 (1960).
\bibitem{hf182}Lee, D.-C., and Halliday, A. N., {\it Nature} {\bf 378},
771 (1995).
\bibitem{wbg}Wasserburg, G. J., Busso, M., and Gallino, R., 
{\it Astrophys. J.} {\bf 466}, L109 (1996).
\bibitem{qw}Qian, Y.-Z., and Wasserburg, G. J., {\it Phys. Rep.} 
{\bf 333--334}, 77 (2000).
\bibitem{qvw1}Qian, Y.-Z., Vogel, P., and Wasserburg, G. J.,
{\it Astrophys. J.} {\bf 494}, 285 (1998).
\bibitem{wq}Wasserburg, G. J., and Qian, Y.-Z., {\it Astrophys. J.} 
{\bf 529}, L21 (2000).
\bibitem{gratton}Gratton, R. G., and Sneden, C., 
{\it Astron. Astrophys.} {\bf 287}, 927 (1994).
\bibitem{mcwill}McWilliam, A., et al.,
{\it Astron. J.} {\bf 109}, 2757 (1995).
\bibitem{sneden1}Sneden, C., et al., {\it Astrophys. J.}
{\bf 467}, 819 (1996).
\bibitem{sneden2}Sneden, C., et al.,
{\it Astrophys. J.} {\bf 496}, 235 (1998).
\bibitem{sneden3}Sneden, C., et al., {\it Astrophys. J.}
{\bf 533}, L139 (2000).
\bibitem{duncan}Duncan, R. C., Shapiro, S. L., and Wasserman, I.,
{\it Astrophys. J.} {\bf 309}, 141 (1986).
\bibitem{qwoo}Qian, Y.-Z., and Woosley, S. E., {\it Astrophys. J.}
{\bf 471}, 331 (1996).
\bibitem{meyer1}Meyer, B. S., et al., {\it Astrophys. J.} {\bf 399},
656 (1992).
\bibitem{janka1}Witti, J., Janka, H.-Th., and Takahashi, K.,
{\it Astron. Astrophys.} {\bf 286}, 841 (1994).
\bibitem{janka2}Takahashi, K., Witti, J., and Janka, H.-Th.,
{\it Astron. Astrophys.} {\bf 286}, 857 (1994).
\bibitem{woo}Woosley, S. E., et al., {\it Astrophys. J.} {\bf 433},
229 (1994).
\bibitem{hoff}Hoffman, R. D., Woosley, S. E., and Qian, Y.-Z.,
{\it Astrophys. J.} {\bf 482}, 951 (1997).
\bibitem{hax1}Qian, Y.-Z., et al.,
{\it Phys. Rev. C} {\bf 55}, 1532 (1997).
\bibitem{hax2}Haxton, W. C., et al.,
{\it Phys. Rev. Lett.} {\bf 78}, 2694 (1997).
\bibitem{meyer2}Meyer, B. S., McLaughlin, G. C., and Fuller, G. M.,
{\it Phys. Rev. C} {\bf 58}, 3696 (1998).
\bibitem{fuller}Caldwell, D. O., Fuller, G. M., and Qian, Y.-Z.,
{\it Phys. Rev. D} {\bf 61}, 123005 (2000).
\bibitem{qian}Qian, Y.-Z., {\it Astrophys. J.} {\bf 534}, L67 (2000).
\bibitem{qvw2}Qian, Y.-Z., Vogel, P., and Wasserburg, G. J.,
{\it Astrophys. J.} {\bf 506}, 868 (1998).
\bibitem{qvw3}Qian, Y.-Z., Vogel, P., and Wasserburg, G. J.,
{\it Astrophys. J.} {\bf 524}, 213 (1999).
\bibitem{israel}Israelian, G., et al., {\it Nature} {\bf 401}, 142 (1999).
\end{references}
\end{document}